\documentstyle[epsfig,12pt]{article}
\newcommand{\doublespace}{
    \renewcommand{\baselinestretch}{1.6}\large\normalsize}

\def\lsim{\mathrel{\rlap{
\lower4pt\hbox{\hskip-3pt$\sim$}}
    \raise1pt\hbox{$<$}}}     
\def\gsim{\mathrel{\rlap{
\lower4pt\hbox{\hskip-3pt$\sim$}}
    \raise1pt\hbox{$>$}}}     

\newcommand{\beq}{\begin{equation}}
\newcommand{\eeq}{\end{equation}}
\newcommand{\be}{\begin{eqnarray}}
\newcommand{\ee}{\end{eqnarray}}

\def\np{Nucl.~Phys.~}
\def\pr{Phys.~Rev.~}
\def\prl{Phys.~Rev.~Lett.~}
\def\pl{Phys.~Lett.~}

\def\der{\mbox{d}}

\begin{document}
\doublespace
\setcounter{equation}{0}
\begin{titlepage}
\pagestyle{empty}
\vspace{1.0cm}
\vspace{1.0cm}
\begin{center}
\begin{Large}
{\bf{The quark condensate in relativistic\\
nucleus-nucleus collisions}}
\end{Large}
\vskip 0.7in
{\large B.~Friman$^{a,b}$, W.~N\"orenberg$^{a,b}$ and
V.D.~Toneev$^{a,c}$}
\vskip 0.15cm
{ {$^a$ Gesellschaft f\"ur Schwerionenforschung (GSI) \\ D-64220
Darmstadt, Germany}\\ {$^b$ Institut f\"ur Kernphysik, Technische
Universit\"at Darmstadt\\ D-64289 Darmstadt, Germany}\\ {$^c$
Bogoliubov Laboratory of Theoretical Physics\\ Joint Institute for
Nuclear Research\\ 141980 Dubna, Russia}\\ }
\end{center}
\vspace{5mm}

\begin{abstract}
We compute the modification of the quark condensate $\langle \bar{q}
q\rangle$ in relativistic nucleus-nucleus collisions and estimate the
4-volume, where the quark condensate is small ($\langle
\bar{q}q\rangle/\langle\bar{q} q\rangle_0\leq$ 0.1--0.3) using
hadron phase-space distributions obtained with the quark-gluon
string model. As a function of the beam energy the 4-volume rises
sharply at a beam energy $E_{lab}/A \simeq$ (2--5) GeV, remains
roughly constant up to beam energies $\simeq 20$ GeV and rises at
higher energies. At low energies the reduction of the condensate is
mainly due to baryons, while at higher energies the rise of the
4-volume is due to the abundant mesons produced. Based on our
results we expect that moderate beam energies on the order of 10
GeV per nucleon are favourable for studying the restoration of
chiral symmetry in a baryon-rich environment in nucleus-nucleus
collisions.
\end{abstract}
\end{titlepage}
\section{Introduction and basic equations}
In  vacuum the chiral symmetry of QCD is spontaneously broken and
the quark condensate $\langle\bar{q} q\rangle$, which is an order
parameter of the chiral phase transition, is non-zero. In hadronic
matter, the quark condensate is reduced, implying a partial
restoration of chiral symmetry, while in quark-gluon matter, beyond
the deconfinement transition, one expects chiral symmetry to be
restored and consequently the quark condensate to vanish. The
leading density and temperature dependence of the condensate has
been studied in static (equilibrium) systems~\cite{GL,CFG}. In this
letter we estimate the modification of the quark condensate in an
inherently dynamic system, namely in a relativistic nucleus-nucleus
collision.

The value of the quark condensate in vacuum will be probed by an
accurate measurement of the low-energy $\pi\pi$ scattering lengths
\cite{DP98}. However, the in-medium modification of the condensate
is, at least presently, not accessible in experiment. Nevertheless,
in the present understanding of hot and dense hadronic matter, the
restoration of chiral symmetry plays a central role. Consequently,
the quark condensate in matter is an essential quantity in the
description of such systems. It has been argued that the in-medium
masses of hadrons are reduced when the chiral symmetry is partially
restored, i.e., when the quark condensate is reduced in magnitude
\cite{BR}. This idea is supported by QCD sum-rule calculations of
vector meson masses in matter \cite{HL}. If this idea is correct,
the observation of changes of hadron masses may open an indirect
way of exploring the restoration of chiral symmetry in dense
matter.

The low-density and low-temperature behaviour of
$\langle\bar{q}q\rangle$ is governed by the relation~\cite{GL,CFG},
\begin{equation}
\label{old}
{\langle\bar{q} q\rangle\over \langle\bar{q} q\rangle_0} =
1 - \sum_h{\sigma_h \rho^s_h\over f_\pi^2 m_\pi^2} \ ,
\end{equation}
where the sum runs over hadron species $h$. Here $\sigma_h$ denotes
the $\sigma$-commutator of the relevant hadron, $\rho^s_h$ the
corresponding scalar density of non-interacting particles, $f_\pi =
94$~MeV the pion decay constant and $m_\pi$ the pion mass.

For nuclear matter at zero temperature and low baryon density
$\rho$, the leading term in Eq.~(\ref{old}) is given by a gas of
non-interacting nucleons. Since, in the rest frame of the system,
the nucleon scalar density approximately equals the baryon density,
one finds~\cite{CFG} using $\sigma_N = 45$ MeV (see
ref.~\cite{sainio})
\begin{equation}
\label{CFG}
{\langle\bar{q} q\rangle\over \langle\bar{q} q\rangle_0} \simeq
1 - {\rho\over 3 \rho_0} \ ,
\end{equation}
which if higher-order  terms are neglected implies that the quark
condensate is reduced by about one third at normal nuclear matter
density $\rho_0 = 0.16\, \mbox{fm}^{-3}$. The terms of higher order
in the density have been estimated in relativistic
Brueckner-Hartree-Fock calculations using realistic nucleon-nucleon
interactions \cite{LiKo,BroWei,Weise}. Qualitatively these
calculations show that the quark condensate is indeed reduced
approximately linearly with density up to $\rho \sim 1.5 \rho_0$.
At somewhat higher densities the condensate seems to stay at a
value of about 40~\% of its vacuum value~\cite{Weise}. We can
account for this effect in an approximate manner by requiring that
the reduction of $\langle\bar{q}q\rangle/\langle\bar{q}q\rangle_0$
due to baryons should not exceed 60$\%$. Eventually, at least at
densities where quark degrees of freedom become relevant, one
expects chiral symmetry to be restored and consequently the
condensate to vanish. However, at beam energies, where such
densities could be reached, the meson densities are large and
dominate. Consequently, the precise behaviour of the modification
of the quark condensate due to baryons at very large baryon
densities is not expected to change our results qualitatively.

At low temperatures and zero net baryon density, the modification
of the quark condensate in Eq.~(\ref{old}) is dominated by
pions. In the chiral limit, where the explicit symmetry breaking
vanishes and $m_\pi \rightarrow 0$, the relative change of the
quark condensate in a pion gas is given
by~\cite{GL}
\begin{equation}
\label{GL}
{\langle\bar{q} q\rangle\over \langle\bar{q} q\rangle_0} =
1 - {T^2\over 8 f_\pi^2}
\end{equation}
to leading order. Since two- and three-loop contributions do not
significantly change the temperature dependence of the quark
condensate \cite{GL,GB2}, we feel that it is justified to neglect
higher-order terms in (\ref{old}) for pions and other mesons.

Consequently, we employ relation (\ref{old}) with a finite pion
mass and a slightly modified form thereof, which accounts for the
higher-order terms in the baryon density as discussed above.
Further details are given in section 3, where the implementation in
the transport calculation is discussed.

While it may be possible to describe the final stages of
nucleus-nucleus collisions approximately with equilibrium
thermodynamics this is clearly not possible for the early stages.
The non-equilibrium character of nucleus-nucleus collisions can be
taken into account by evaluating the scalar densities in Eq.
(\ref{old}) with the corresponding, possibly non-equilibrium,
phase-space densities, obtained using a model for the space-time
evolution of nucleus-nucleus collisions. Before we turn to this
problem, we present some qualitative arguments for the behaviour of
the quark condensate in heavy-ion collisions.

\section{Qualitative consideration of heavy-ion collisions}

In contrast to  equilibrium systems, where the quark condensate has
been studied so far, nucleus-nucleus collisions are characterized
by a short time scale on the order of 10~fm/c.  So far the typical
time scale $\tau$ for the quark condensate to respond to changes of
the medium has  not been studied. However, one would expect this
time scale to be governed by the mass of the lightest scalar meson
and thus to be of the same order as the hadronic time scales, i.e.
$\tau \lsim 1$ fm/c. In this exploratory calculation we assume that
the response of the condensate is instantaneous, except for
produced particles, whose contribution to the quark condensate is
taken into account only after a formation (proper) time $\tau_f =$
1~fm/c.

Before we turn to nucleus-nucleus collisions, let us consider the
quark condensate in a moving nucleus. Since the quark condensate is
a Lorentz scalar, it must be independent of the reference frame of
the observer. To illustrate this, we repeat the lowest-order
calculation \cite{CFG}, which leads to Eq.~(\ref{old}). The leading
contribution to the energy density of a system of nucleons is given
by
\begin{equation}
\label{eps}
\varepsilon = \int \der^3p \ \sqrt{\vec{p}^2 + m_N^2} \
n_{\vec{p}} \ ,
\end{equation}
where $n_{\vec{p}}$ is the nucleon distribution function.
The modification of the quark condensate due to the presence of
the nucleons is, according to the Feynman-Hellmann theorem,
\begin{equation}
\label{moving}
\langle\bar{q}q\rangle - \langle\bar{q}q\rangle_0 = \frac{d
\varepsilon}{d m_q} = \frac{\sigma_N}{m_q}
\int \der^3p \ \frac{m_N}{\sqrt{\vec{p}^2 + m_N^2}} \ n_{\vec{p}} \ ,
\end{equation}
where $\sigma_N = m_q{d m_N}/{d m_q}$. The integral on the
right-hand side of (\ref{moving}) is the scalar nucleon density.
Since $\der^3p/\sqrt{\vec{p}^2 + m_N^2}$ and $n_{\vec{p}}$ are
Lorentz invariants, the value of the scalar density and
consequently the quark condensate are independent of the reference
frame~\footnote{We note that these arguments are not restricted to
equilibrium systems; the distribution function $n_{\vec{p}}$ in
(\ref{eps}) and (\ref{moving}) is arbitrary.}.

It follows that  the condensate in the interior of a moving nucleus
equals the condensate in a nucleus at rest, i.e., if the
lowest-order approximation (\ref{CFG}) is valid, it is reduced by
about 1/3 from its vacuum value. This means that when two nuclei
pass through each other without interacting, the quark condensate
in the overlap region is reduced by the amount corresponding to
nuclear matter at $\rho = 2 \rho_0$, i.e., by about 2/3. On the
other hand, the baryon density in a moving system is enhanced by
Lorentz contraction of the volume $\rho = 2\gamma \rho_0$, where
$\gamma = 1/\sqrt{1-v^2}$ and $v$ is the  velocity of the nucleus.
At AGS energies this purely kinematical effect without any
dynamical compression yields a baryon density of (5-6)$\rho_0$,
while at CERN energies one finds about $20 \rho_0$. Similarly, also
the energy density is strongly enhanced partly due to the Lorentz
contraction of the collision volume. This trivial Lorentz
contraction, which is responsible for a substantial fraction of the
very large baryon and energy densities found in simulations of
heavy-ion collisions at ultra-relativistic energies, does not
affect the quark condensate. Consequently, the baryon density is
not a suitable measure for the restoration of chiral symmetry in
such collisions.

When interactions are taken into account, the nucleons are slowed
down and other hadrons (mesons and baryon-antibaryon pairs) are
produced. Both effects reduce the quark condensate. In order to
illustrate this  we consider the Landau model for hadronic
collisions~\cite{Landau}. If, as assumed in this model, the
deceleration is very fast, the interaction volume is not expected
to differ appreciably from the Lorentz contracted collision volume.
Thus, after stopping there is a large density $(\approx
2\gamma\rho_0 )$ of relatively slow hadrons in this volume, which
can lead to a considerable reduction of the quark condensate. For
the stopped nucleons we have $m_N/\sqrt{\vec{p}^2+m_N^2}\sim 1$, so
that their contribution to the scalar density approximately equals
the corresponding part of the baryon density
(cf.~Eq.~(\ref{moving})). Similarly, for the bulk of the produced
hadrons, the velocities are relatively small, so that the
associated scalar densities are well approximated by the particle
densities. This simple consideration shows, that the value of the
quark condensate in nucleus-nucleus collisions depends sensitively
on the collision dynamics, in particular on stopping and particle
production rates.

We note that when  the finite sizes of the hadrons are taken
seriously, one arrives at similar conclusions.  The relevant
quantity, which determines the quark condensate of such a system is
not the number density but the volume fraction which is not
occupied by hadrons \cite{Weise2}. In the spirit of the chiral bag
model it is assumed that the interior of a nucleon is in the
chirally symmetric phase, and consequently that the quark
condensate there is effectively zero. Thus, each nucleon represents
a small volume, where the chiral symmetry is locally restored, and
the average value of the condensate in nuclear matter is
proportional to the fraction of the volume which is unoccupied. By
identifying the coefficient of the term linear in density with that
given by Eq.~(\ref{old}), one finds for the volume of a nucleon at
rest $v_N=\sigma_N/f_\pi^2m_\pi^2$,  which implies $R_N=0.8$ fm.
Now consider a moving nucleus composed of finite-size nucleons.
Since all volumes, that of the nucleus and those of the nucleons,
are Lorentz contracted by the same factor~\footnote{In this
argument we assume that the internal velocities $v_{i}$ of the
nucleons in the nucleus are small compared to the velocity  $v$ of
the nucleus, so that terms of order $(v_{i}/v)^2$ are negligible.},
the fraction of the volume which is not occupied is a Lorentz
invariant. On the other hand, a nucleon, which is stopped in a
nucleus-nucleus collision, recovers its rest volume. Consequently,
stopping leads to an increase in the occupied fraction of the
Lorentz contracted interaction volume and thus to a decrease of
$|\langle\bar{q}q\rangle|$. The occupied fraction is increased
further by the produced hadrons. Thus, this simple picture provides
an intuitive interpretation of the features discussed above.

\section{The quark condensate in Au on Au collisions}

From the arguments presented in Section~2 it should be clear that a
realistic estimate of the quark condensate in relativistic
nucleus-nucleus collisions can be obtained only by using a reliable
model for the space-time evolution of the collision. We employ the
Quark-Gluon String Model (QGSM)~\cite{TA90}, which is based on the
string phenomenology of hadronic interactions. Baryons and mesons
belonging to the two lowest $SU(3)$ multiplets along with their
antiparticles are included. The interactions between the hadrons
are described by a collision term, where the Pauli principle is
imposed in the final states. This includes elastic collisions as
well as hadron production and decay processes.  As mentioned above,
a proper time $\tau_f = 1$~fm/c for the formation of hadrons is
incorporated. Mean fields and interactions between strings are not
taken into account in the present version of the model. The general
experimental characteristics of relativistic nucleus-nucleus
collisions are well reproduced over a large range of beam energies
from SIS \cite{SIS}, to AGS \cite{AGS} and SPS energies
\cite{TA90,SPS}. Thus, we expect that the model describes the
space-time evolution of such collisions reasonably well.

In Fig.~1 we show the time evolution of the central baryon density
for head-on Au+Au collisions at beam energies from $E_{lab}/A$ = 2
GeV up to 50 GeV. The initial time (t=0) corresponds to the instant
when the two Lorentz-contracted nuclei touch.  The densities are
obtained from the number of particles in the  cylindrical test
volume of radius $R_f$ and length $2R_f/\gamma_{cm} \ $  at rest in
the center-of-mass frame. The factor $1/\gamma_{cm}$ in the
longitudinal direction accounts for the Lorentz contraction of the
colliding nuclei. The total baryon density in the test volume
increases rapidly reaching a maximum shortly after the point of
maximum overlap. Subsequently the density decreases rapidly with
time. We also show the density of participant nucleons, defined as
those that have suffered at least one collision. The maximum baryon
density reached at $E_{lab}/A=10$~GeV is in good agreement with
that found in ref.~\cite{LK95}. The maximum density is reached at
the same time, $t \approx 4$~fm/c, while our maximum value is
somewhat smaller due to a larger test volume; we use $R_f=5$~fm
while in \cite{LK95} a sphere of radius $2$~fm was employed.

A comparison of the total and participant densities shows that even
near the maximum not all nucleons have experienced a collision.
Nevertheless, both definitions of the density imply that very large
baryon densities are reached at high beam energies. However, as we
argued above, these densities are not immediately relevant for the
restoration of chiral symmetry and associated medium effects.

We compute the quark condensate in a given volume $V$ by
implementing Eq.~(\ref{old}) in the transport model in the form
\begin{equation}
\label{qcsim}
\frac{\langle\bar{q}q\rangle}{\langle\bar{q}q\rangle_0} = 1 -
\frac{1}{V}\sum_{i=1}^{N_V} \frac{m_i}{\epsilon_i}\,\frac{\sigma_i}
{f_\pi^2 m_\pi^2} \ ,
\end{equation}
where the sum runs over all particles $N_V$ in $V$. Here
$\sigma_i$, $m_i$ and $\epsilon_i = \sqrt{\vec{p}_i^2 + m_i^2}$
denote the sigma term, mass and energy of particle $i$. For nucleons
and pions we use $\sigma_N = 45$ MeV and $\sigma_\pi = m_\pi/2$,
respectively, while for all other hadrons we take the sigma term to
be given by $\sigma_i = (Q_i/Q_N)\sigma_N$, where $Q_i$ denotes the
light-valence-quark content of hadron $i$. Produced particles are
counted only after their formation time $(\epsilon_i/m_i)\tau_f$
has passed. As mentioned above we  approximate higher-order
effects by limiting the modification of the quark condensate due to
baryons to be less than 0.6, cf.~\cite{BroWei,Weise,Weise2}.

The resulting time evolution of the quark condensate is shown in
Fig.~2 for the same test volume as for the baryon densities in
Fig.~1. For beam energies beyond 4 GeV, the ratio $\langle\bar{q}q
\rangle/\langle\bar{q}q\rangle_0$ becomes negative for some time
during the collision. Clearly, the low-density approximation is no
longer valid when this happens. However, if we are interested only in
the time $\Delta t_{qc}$ that the system spends in a state where the
quark condensate is small, say
$\langle\bar{q}q\rangle/\langle\bar{q}q\rangle_0\leq 0.3$,
this approximation is expected to be reasonable. Indeed,
in this case only the reliability of the approximation for
$\langle\bar{q}q\rangle/\langle\bar{q}q\rangle_0 > 0.3$ matters. We
note at this point that possible modifications of the dynamics,
e.g. due to an equation of state with a strong first-order phase
transition, are not included in the transport model. However, such
effects are not expected to change the expansion time scale
dramatically~\cite{FKR}.

At a beam energy of $E_{lab}/A=2$~GeV the pion density remains
relatively small, and hence the main reduction of
$|\langle\bar{q}q\rangle|$ is due to baryons which in our model
reach densities of $3\rho_0$ (see Fig.~1). At higher energies, the
pion contribution becomes gradually more important. At $E_{lab}/A =
50$ GeV and above the reduction of $|\langle\bar{q}q\rangle|$ is
dominated by pions and other mesons. We note that the time period
$\Delta t_{qc}$, where $\langle\bar{q} q\rangle / \langle\bar{q}
q\rangle_0 < r$, with $r = 0.1 - 0.3$, in the cylindrical test
volume is between 5 and 7 fm/c for all bombarding energies beyond 5
GeV. At SIS energies and partially also at AGS energies, the
reduction of the quark condensate is induced predominantly by the
baryons, while at ultra-relativistic energies the meson
contribution prevails.

In general we expect that observable effects due to partial
restoration of chiral symmetry should depend not only on the time,
where the quark condensate is small, but also on the corresponding
volume. Clearly, in order to have a clean signal one would like the
time to be as long as possible, and in order to minimize unwanted
surface effects the volume should be as large as possible. However,
the relative importance of time and volume probably depends on the
particular signal one considers and is not known in general.
Consequently, an optimal criterion of universal validity is
difficult if not impossible to construct. We choose a relatively
simple quantity that accounts for both time and volume, the
4-volume
\begin{eqnarray}
\Omega_r &=& \int \der^3x\, \der t\,\, \theta \left(r-\frac{\langle \bar{q}q
\rangle(x,t)}
{\langle \bar{q}q\rangle_0}\right)\\&\simeq&
\int\der t\left[\sum_{i}\Delta V_i\,\,\theta \left(r-\frac {\langle
\bar{q}q\rangle_{i}(t)}{\langle
\bar{q}q\rangle_0}\right)\right],
\end{eqnarray}
where $\theta$ denotes the Heaviside function. In the second line
we indicate how $\Omega_r$ is computed in the transport model.
Using Eq.~\ref{qcsim} we evaluate the quark condensate
$\langle\bar{q}q\rangle_{i}(t) $ in each cell $i$ at a given time
t, and then sum the volumes $\Delta V_i$ of those cells where
$\langle\bar{q}q\rangle/\langle\bar{q}q\rangle_0$ is smaller than
$r$. Finally we integrate over time. Thus the 4-volume, where the
quark condensate is small, is determined dynamically. We make use
of the cylindrical symmetry of central collisions to reduce the
number of cells. The cells are rings around the collision axis of
longitudinal and radial extension $\Delta z=2$ fm and $\Delta r =
1$ fm, respectively.

The 4-volume $\Omega_r$ is shown by the full lines in Fig.~3 as a
function of the bombarding energy for different values of $r$. Here
we employ the modified version of Eq.~(\ref{old}), where the baryon
contribution to the quark condensate is limited. We find a sharp
increase at a threshold energy of a few GeV per nucleon, a plateau
in an intermediate energy range up to $\sim 20$ GeV per nucleon and
an increasing 4-volume for energies beyond that. The threshold and
the plateau regions are associated with high baryon density, while
the rise at high energies is due largely to meson degrees of
freedom. At this point we note that the saturation of the baryon
contribution, introduced to account for higher-order effects, plays
a crucial role at low energies but does not affect the 4-volume
appreciably at high beam energies, $E_{lab} \geq 3$ GeV.

In order to obtain a quantitative characterization of the different
regimes we also show the 4-volume, where the quark condensate is
small with the additional constraint that the baryon scalar density
should be larger than $1.5\rho_0$ (dashed lines in Fig.~3). This
4-volume exhibits a maximum at a relatively low beam energy, and
then decreases as the energy is increased. The reduction of the
contribution from baryon-rich matter is basically due to the
shortening of the corresponding time scale (Fig.~2). This in turn
is a result of the higher expansion velocity, which leads to a
faster dilution of the baryon density (cf. Fig.~1). Since the net
baryon number is constrained by conservation laws, a more rapid
expansion cannot be compensated for by an enhanced rate for
particle production. On the other hand, the growth of the meson
contribution to the 4-volume at high energies is due to a high rate
for meson production, which compensates for the strong expansion
and implies that a dense meson gas exists over a relatively long
time scale in an expanding interaction volume.

The characteristics of the restoration of chiral symmetry at high
baryon density may differ from that at high temperature. At zero
baryon density and finite temperatures, the pion is the most
abundant hadron. The thermal pions induce a modification of the
quark condensate to lowest order, which in the chiral limit gives
rise to the well known $T^2$ term. However, because the pion is a
Goldstone boson, and consequently its interaction terms involve
gradients, the leading correction to e.g. the mass of the $\rho$
meson in a pion gas \cite{Ioffe} is ${\cal O}(T^4)$. Thus, the
modification of meson masses as a function of temperature is not
simply related to the quark condensate. On the other hand, at zero
temperature and finite baryon density, the leading corrections to
the quark condensate and hadronic masses are due to nucleons. Here
the situation may be different, since e.g. non-gradient
$\rho$-nucleon couplings are allowed \footnote{See however
ref.~\cite{Birse} for a critical discussion of this point.}.

At this point we also note that the in-medium modifications invoked
in interpretations of the low-mass lepton-pairs in
ultra-relativistic nucleus-nucleus collisions at the
SPS~\cite{CERES,HELIOS} depend mainly on high baryon densities
\cite{LKB,RW,CBRW}. For such medium effects the dashed curves in
Fig.~3 and the baryon time scales of Fig.~2 are relevant.
Consequently, if such a model is correct, one expects an even more
pronounced enhancement of lepton pairs at lower beam energies, say
$E_{lab}/A\sim 10$ GeV, where high baryon densities dominate.

In the baryon-rich case, the 4-volume exhibits a maximum, which --
depending on the value of $r$ -- lies in the range 2--8~GeV. In
principle this indicates the optimal beam energy for exploring
effects connected with the restoration of chiral symmetry
associated with high baryon densities in heavy-ion collisions.
However, the quark condensate is not a direct observable.
Furthermore, as noted above, the relative weight of the time and
volume depends on the probe under consideration. Consequently,
there is at best an indirect connection between the maximum in the
4-volume and the optimal energy for a given probe.

However, the threshold behaviour of the 4-volume is a
characteristic feature, which we expect to be of general validity
for any probe of the restoration of chiral symmetry in dense and
hot matter. The threshold energy for a given observable may differ
from our estimate $E_{lab}/A \simeq (2-4)$ GeV, due to the unknown
relative importance of the time and volume and additional
uncertainties, like e.g. the response time of a given probe to
changes in the quark condensate. Nevertheless, our results indicate
that one should expect significant effects due to partial
restoration of chiral symmetry in a large volume over an extended
period of time already at moderate beam energies, say 2 GeV $<
E_{lab}/A < 10$ GeV.

\section{Conclusions}

We argue that the quark condensate is better suited than the baryon
density as a measure for chiral restoration in simulations of
relativistic nucleus-nucleus collisions. The condensate is directly
related to the spontaneous breaking of chiral symmetry, since it is
an order parameter for this transition. Furthermore, it is not
encumbered by trivial Lorentz contraction effects and accounts for
both baryons and mesons in a natural way.

We have estimated the quark condensate in relativistic
nucleus-nucleus collisions. We find that the invariant 4-volume,
where the condensate is small, increases rapidly at $E_{lab}/A =
(2-5)$ GeV, levels off at intermediate energies and increases for
beam energies beyond $\sim 20$ GeV. This behaviour is due to an
interplay between meson and baryon contributions to the quark
condensate. For baryon-rich matter, the 4-volume decreases at high
energies with increasing beam energy, giving rise to a maximum at
fairly low beam energies. Although for a given probe the optimal
beam energy may vary, experimental signatures of chiral symmetry
restoration are expected to exhibit a characteristic threshold
behavior corresponding to the sharp rise of the 4-volume. All in
all, our results indicate that the conditions reached in
nucleus-nucleus collisions at moderate beam energies, 2~GeV $<
E_{lab}/A < 10$~GeV, are favourable for exploring the restoration
of chiral symmetry in dense/hot matter. For probes that rely on
high baryon densities, the optimal conditions are probably reached
at beam energies around $E_{lab}/A \sim 10$ GeV.

We thank H.J.~Pirner and W.~Weise for useful discussions and
W.~Cassing for pointing out the importance of a dynamical treatment
of the meson dominated regime at high energies. V.D.T. gratefully
acknowledges the hospitality of GSI, where this work was done, as
well as the partial support by the WTZ-program of BMBF.

\newpage
\setlength{\unitlength}{1mm}
\begin{picture}(150,170)
\put(0,0){\epsfig{file=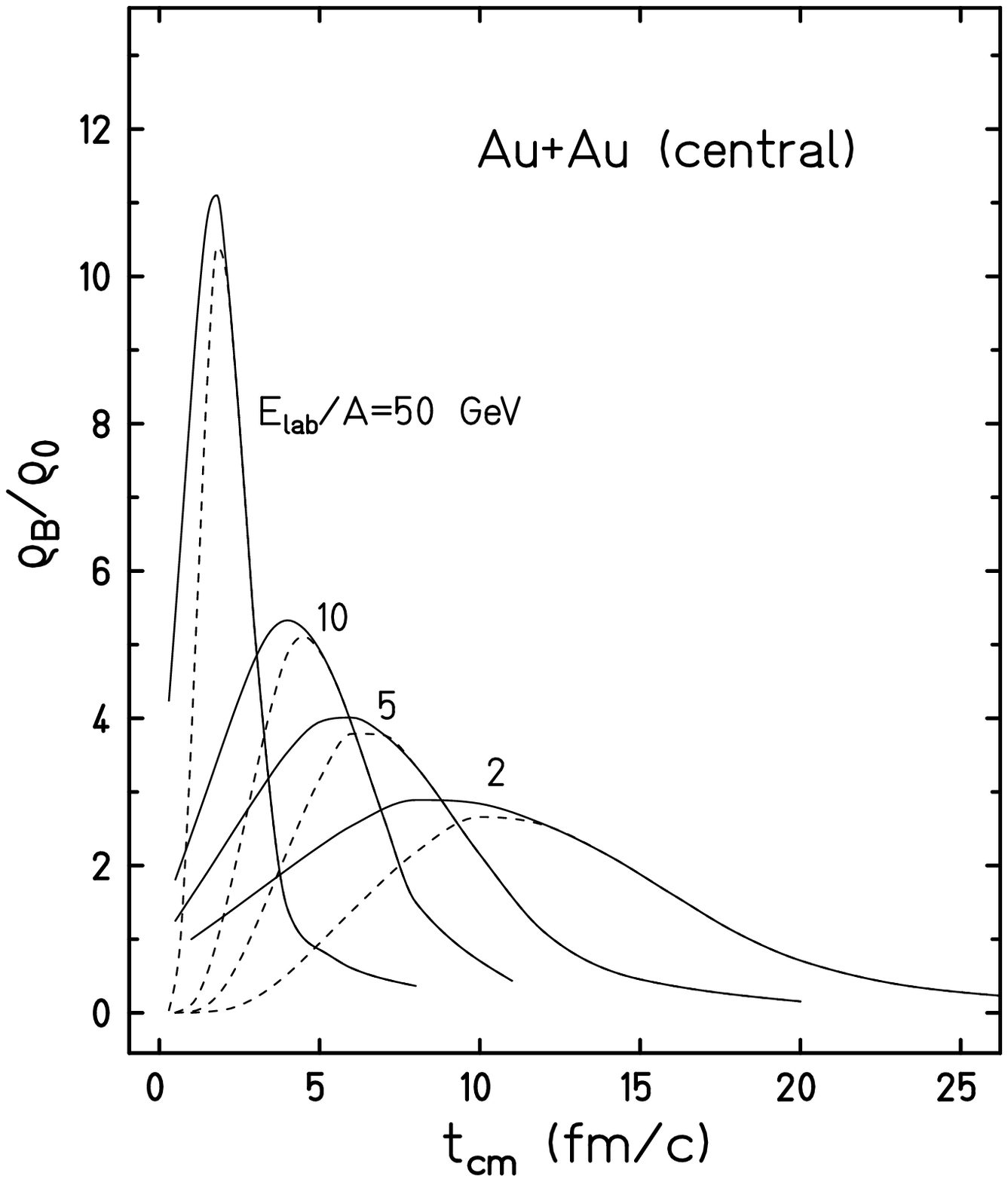,height=160mm}}
\end{picture}

Fig.~1. Time evolution of the central baryon density in head-on Au+Au
collisions for beam energies 2~GeV $\le E_{lab}/A \le$ 50~GeV.
The solid  lines correspond to all baryons while the dashed lines
are for the participants only.

\begin{picture}(150,160)
\put(0,0){\epsfig{file=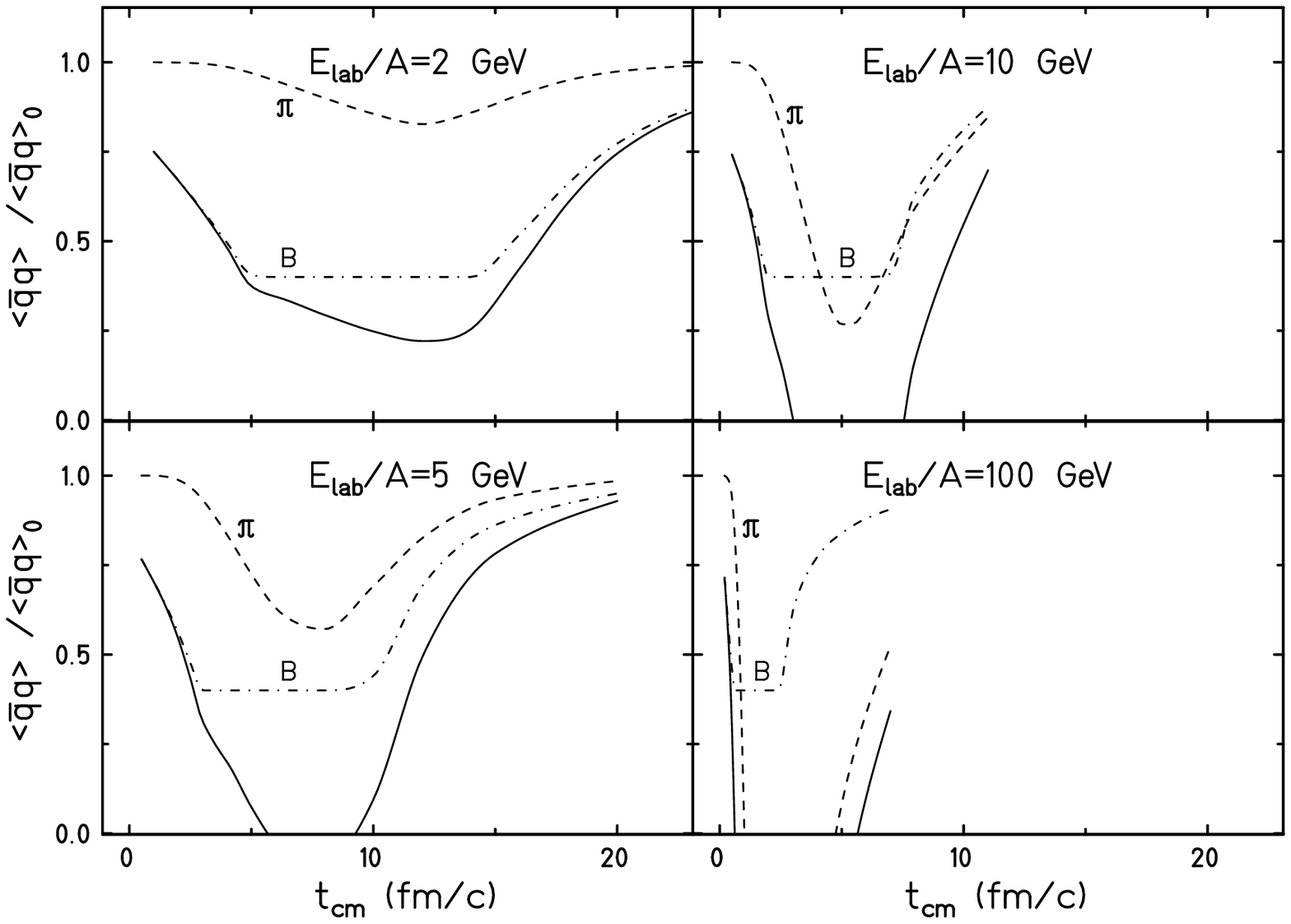,width=150mm}}
\end{picture}

Fig.~2. Time evolution of the central quark condensate (solid lines)
in head-on Au+Au collisions for $E_{lab}/A=$ 2, 5, 10, 100~GeV.
The contribution from baryons (dash-dotted lines) and pions
(dotted  lines) are shown separately.

\begin{picture}(150,170)
\put(0,0){\epsfig{file=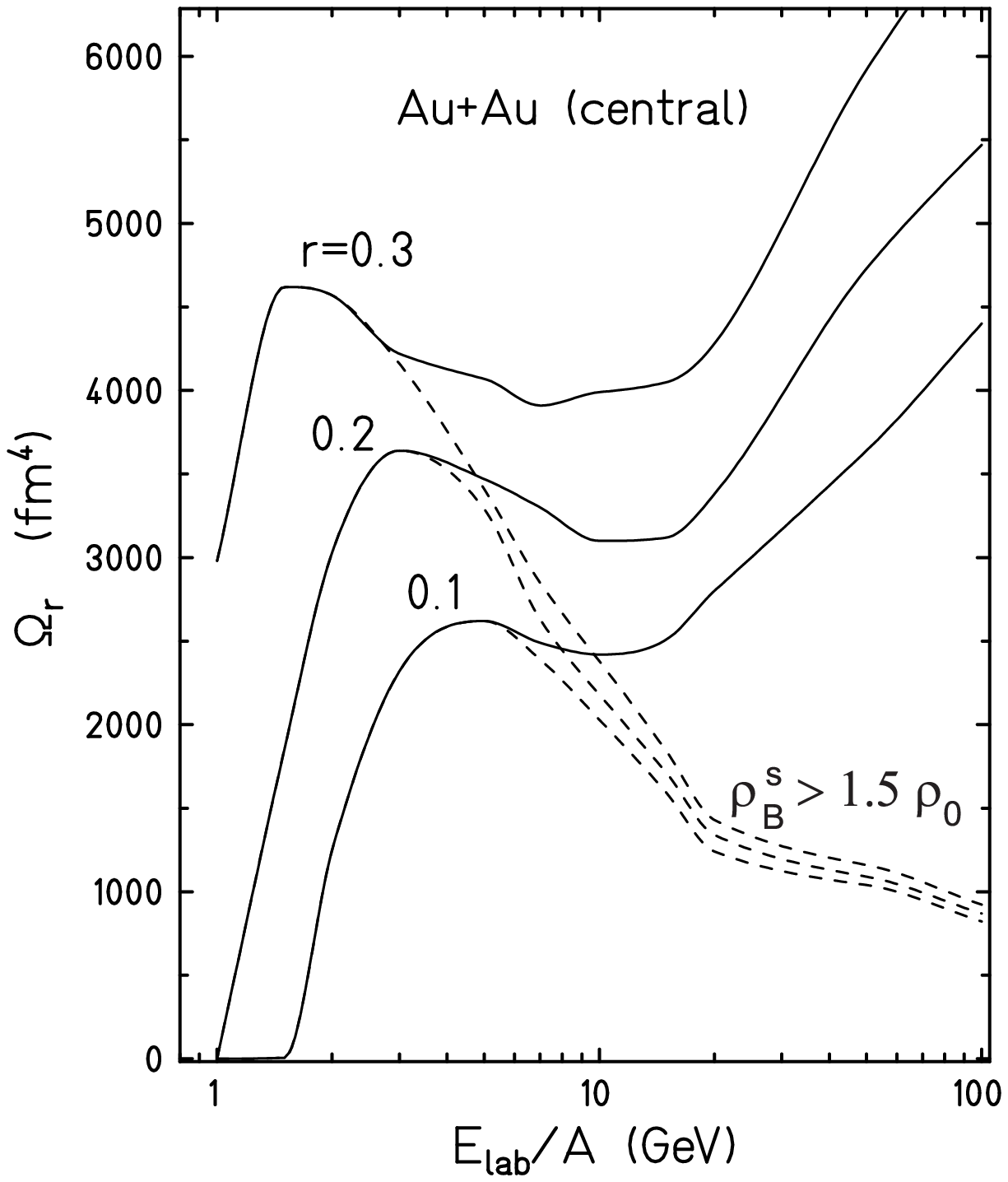,height=160mm}}
\end{picture}

Fig.~3. The invariant 4-volume $\Omega_r$ (solid line), where
$\langle \bar{q} q\rangle/\langle \bar{q} q\rangle_0$ drops below
$r= 0.1, 0.2$ and $0.3$, as function of the beam energy. The
corresponding 4-volume, where in addition the scalar baryon density
exceeds 1.5 $\rho_0$, is shown by the dashed lines.

\end{document}